\documentstyle[aps,preprint,epsf]{revtex}
\tightenlines
\begin{document}
\title{The Potential for the 
Phase of the Wilson line at Nonzero Quark Density}
\author{C. P. Korthals Altes$^a$,
Robert D. Pisarski$^b$, and Annamaria Sinkovics$^c$}
\address{
$^a$Centre Physique Theorique au CNRS,\\
Case 907, Luminy, F13288, Marseille, France\\
email: Chris.Korthal-Altes@cpt.univ-mrs.fr\\
$^b$Department of Physics,\\ Brookhaven National Laboratory, 
Upton, New York 11973\\
email: pisarski@bnl.gov\\
$^c$The Randall Laboratory of Physics\\ 
University of Michigan,
Ann Arbor, MI 48109-1120\\
email: ani@umich.edu}
\maketitle

\begin{abstract}
\begin{quotation}
The contribution of quarks to the
effective potential for the phase of the Wilson line is computed
at nonzero temperature and quark chemical potential to one 
and two loop order.  
At zero temperature, regardless of the
value of the quark chemical potential, 
the effective potential for the phase of the Wilson line vanishes.
At nonzero temperature, 
for special values of the phase the free energy of quarks
equals that of bosons; at nonzero
chemical potential, such quarks can ``Bose condense'',
albeit with negative density.
\end{quotation}
\end{abstract}

At nonzero temperature, $SU(N)$ gauge theories in the absence
of dynamical fermions 
have a deconfining phase transition related to the spontaneous
breaking of a global $Z(N)$ symmetry \cite{zn}.  The order parameter 
for this transition is the Wilson line:
its vacuum expectation value 
vanishes in the confined, low temperature phase,
and is nonzero in the deconfined, high temperature phase.
Because of the global $Z(N)$ symmetry, in the deconfined
phase there are $N$ equivalent vacua, 
with the phase of the Wilson line equal to one of the $N^{th}$ roots
of unity \cite{first}-$\!\!$\cite{belyaev}.  
The interface tension between these $N$ vacua is computable
semiclassically \cite{interf}-$\!\!$\cite{anti}, by computing
the effective potential for the phase of the Wilson line.
This interface tension is measurable from 
numerical simulations on the lattice \cite{lattice}.
In the presence of dynamical quarks in the fundamental
representation, the Wilson line acquires a vacuum
expectation value at all temperatures, and the vacuum is unique,
with zero phase for the Wilson line.
Nevertheless, depending upon the number of colors and quark flavors,
metastable states in which the phase of the Wilson
line is nonzero can arise \cite{misc}, and may be of physical 
interest (although this is controversial \cite{anti}).

The effective potential for the phase of the Wilson line
for an $SU(N)$ gauge theory has been computed at one \cite{first}
and two \cite{twoloop}-$\!\!$\cite{altes} loop
order at a nonzero temperature, $T$.  Here we extend this result
to a nonzero quark chemical potential, $\mu $.

The Wilson line is
\begin{equation}
\Omega(\vec{x}) = exp\left(i g \int^{\beta}_0 A_0(\vec{x},\tau) 
d \tau \right) \; ,
\end{equation}
where $T$ is the temperature, $g$ the gauge coupling constant,
and the imaginary time $\tau : 0 \rightarrow \beta = 1/T$.
To represent a nonzero phase of the Wilson line in the deconfined
phase, we can take a constant background 
value for the time-like component of the 
gauge potential, $A_0$ \cite{first}-$\!\!$\cite{anti}.  
After a global color rotation, this can be chosen to be 
a diagonal matrix:
\begin{equation}\label{gen}
A_0 = \frac{2 \pi T}{g} 
\left[\begin{array}{cccc}
q_1 & 0 & \ldots & 0\\
0 & q_2 & \ldots & 0 \\
0 & 0 & \ddots & 0 \\
0 & 0 & \ldots & q_N 
\end{array}\right]\; , \;\;
\sum^N_{j = 1} q_j = 0 \; .\quad 
\end{equation}
The sum of the $q_j$ must vanish so that $A_0$ is an
$SU(N)$, and not a $U(N)$, matrix.  
For example, consider $Z(N)$ degenerate vacua \cite{interf}
in the pure glue theory.
In the trivial vacuum the phase of the Wilson line vanishes,
$\Omega = {\bf 1}$,
so all $q_j =0$.  The equivalent vacuum in which the
Wilson line is the first $N^{th}$ root of unity, 
$\Omega = exp(2 \pi i/N) {\bf 1}$, is given by 
$q_1 = \ldots q_{N-1} = q/N$, and $q_N = - (N-1) q/N$;
tunneling between these two vacuum is parametrized by taking
$q:0\rightarrow 1$.

Because a constant $A_0$ field can be eliminated by a global
gauge transformation,
the classical action is independent of the value of $A_0$.
At nonzero temperature, 
this degeneracy is lifted by quantum effects,
as the required gauge transformation alters the boundary conditions
required for the fields in imaginary time.  Consequently,
at one loop order and beyond, a potential for $A_0$, 
or more properly for the phase of the Wilson line, is generated.  
In terms of partition functions \cite{roberge}, this potential
is generated because one is summing
over states which differ by a global gauge transformation.

We compute the quark contribution to the effective action for a single,
massless flavor.  At one loop order, this is just the free energy in the
presence of a background $A_0$ field:
\begin{equation}
{\cal S}^1 = -Tr \log{ i \left[ D\kern-0.22cm /^{cl} - 
\mu \gamma_0 \right] } \; ;
\end{equation}
$D\kern-0.22cm /^{cl} = \partial\kern-0.19cm / - i g 
A\kern-0.18cm /$ is the covariant derivative,
and the trace includes color, Dirac indices, and a loop integral.
We include the effects of a Fermi sea for quarks by introducing
a chemical potential for baryon number, $\mu $.

In the following it helps to introduce two types of variables
related to $\mu $ and the $q_j$.
One set of variables are $N$ color-dependent ``chemical potentials'':
\begin{equation}\label{mudef}
\mu _j = \mu + 2 \pi T i \, q_j \; \; , \;\; j=1\ldots N \; .
\end{equation}
Notice that the $q_j$ act like an 
imaginary chemical potential for color \cite{roberge}.
Since the $\mu _j$ are complex, the effective action is as well.

Alternately, we also use the variables $C_j$,
\begin{equation}\label{defc}
C_j = q_j + \frac{1}{2} + i \frac\mu {2 \pi T} \; \; , \;\; j=1\ldots N \; .
\end{equation}
The $\mu _j$ are useful in comparing to the limit
in which $\mu \neq 0$ and $A_0 \rightarrow 0$; 
the $C_j$ are convenient in the
limit that  $A_0 \neq 0$ and $\mu \rightarrow 0$.

As the background gauge field is diagonal in color, at one loop
order the color trace is trivial,
\begin{equation}
{\cal S}^1  = \sum^N_{j=1} {\cal S}_j =
- \sum^N_{j=1} \log{  \left[ i \partial\kern-0.19cm / - i 
\mu _j \gamma_0 \right]} \; .
\end{equation}

We compute the ${\cal S}_j$ in two different ways.  The first is careful and
certainly correct, but awkward.
The trace in ${\cal S}_j$
involves an integral over the loop momentum, 
$K^\mu = (k^0,\vec{k})$, where for 
fermions in imaginary time, the energies are discrete, 
$k_0 = (2 l + 1) \pi T$ for integral $l = \pm 1, \pm 3 \ldots$.  
Performing the sum over $l$ by contour integration, 
\begin{equation}\label{contour}
{\cal S}_j = - \beta V \; 2 T^3
\int \frac{d^3 k}{(2 \pi)^3} \left\{ \log \left[ 1 + e^{- 
(k - \mu _j)/T } \right] +  \log \left[ 1 + e^{- (k + \mu _j)/T }\right]
\right\} \; ;
\end{equation}
$V$ is the volume of space, so 
$\beta V$ is the volume of space-time.
To do the integral over spatial momentum, we use
zeta function regularization; 
this was used at $T\neq 0$ by Actor \cite{actor1,actor2} and by
Weldon \cite{weldon}.  
Expanding the logarithm and integrating over $\vec{k}$ gives 
\begin{equation}
{\cal S}_j = - \beta V \;
\frac{4 T^4}{\pi^2} \sum_{m=1}^{\infty} (-1)^{m+1} 
\frac{1}{m^4} \cosh{\left( \frac{m \mu _j}{T}\right)} \; .
\end{equation}
In each $\cosh$ function we expand as a power series, 
interchange the order of the two series, and so obtain a sum over
$\eta$ functions, 
\begin{equation}\label{eta}
{\cal S}_j = - \; \beta V \;
\frac{4 T^4 }{\pi^2} \sum_{n=0}^{\infty} \frac{1}{(2 n)!} 
\left(\frac{\mu _j}{T}\right)^{2n }\eta(4-2n) \; ,
\end{equation}
where the $\eta$ function is defined as
\begin{equation}
\eta(s) = \sum_{i=1}^{\infty} \left(-1 \right)^{i+1} \frac{1}{i^s}=
(1-2^{1-s}) \zeta(s)~~~ Re(s)>1 \; .
\end{equation}
Zeta function regularization is used to define the sum over negative $s$.
Then a wonderful simplication enters: $\eta(-2m) = 0$ for any
integral $m$, so only three terms contribute to ${\cal S}_j$,
\begin{equation}
{\cal S}_j = - \; \beta V \;
\frac{4 T^4}{\pi^2} \left\{ \eta(4) + \frac{\eta(2)}{2} 
\left(\frac{\mu _j}{T} \right)^2
+ \frac{\eta(0)}{4!} \left(\frac{\mu _j}{T}\right)^4 \right\} 
\end{equation}
Using the explicit values of the $\eta$ function, 
\begin{equation}\label{oneloopmu}
{\cal S}^1 =  - \; \beta V \; \left\{ 
\frac{7\pi^2}{180}\; N T^4 + \frac{T^2}{6} 
\sum_{j=1}^N \mu _j^2 + 
\frac{1}{12 \pi^2} \sum_{j=1}^N \mu _j^4  \right\} \; .
\end{equation}
For each $\mu _j$, this expression is periodic 
in $q_j$ modulo $1$ over the interval $\in (-1/2, 1/2)$.
The real part of ${\cal S}^1$ agrees with Actor \cite{actor1};
he neglected the imaginary part.

The second way to compute ${\cal S}^1$ is just to 
take the result for $\mu = 0$, and substitute the $C_j$ from (\ref{defc}):
\begin{equation}\label{oneloopc}
{\cal S}^1=- \; \beta V \; {4\over 3} \pi^2 T^4 
\left( \sum_{j=1}^N B_4(C_j) - \frac{N}{16}\right) \; ,
\label{oneloop}
\end{equation}
where $B_4(x)$ is the Bernoulli function,
\begin{equation}
B_4(x)= x^2 (1-x)^2 \; ;
\end{equation}
in this expression and henceforth, the real part of $x$ is defined
modulo 1.
It can be shown that (\ref{oneloopmu}) and (\ref{oneloopc}) agree.
When the $q_j=0$, it is evident that (\ref{oneloopmu}) reduces
to the free energy for a massless fermion at $\mu $ and $T\neq 0$
\cite{kapusta}.
When $\mu = 0$, (\ref{oneloopc}) obviously reduces to the usual
effective potential for the $q_i$ at $T \neq 0$.

While the first derivation was cumbersome, it has the virtue that 
in (\ref{eta}) we
only need the $\eta(s)$ function for negative values of $s$.
In contrast, with the second derivation it is necessary
to analytically continue the Bernoulli function $B_4(x)$ for complex
values of $x$, which is valid for $|x-1/2| \leq 1$.
Both (\ref{oneloopmu}) and (\ref{oneloopc})
are analytic continuations of the
same function, (\ref{contour}),
at nonzero values of the chemical potential.

Given our success at one loop order, however,
we can follow the same process at two loop order.  In other
words, in order to compute the effective potential at two
loop order, we just take the result for 
$\mu =0$ \cite{twoloop}-$\!\!$\cite{altes}, and substitute
the values of $C_j$ in (\ref{defc}).  This gives
$$
{\cal S}^2=- \; \beta V \; {g^2\over4} T^4 \left[\sum_{j \neq k = 1}^N 
B_2(C_{jk}) \left( B_2(C_j)+B_2(C_k) \right)-B_2(C_j)B_2(C_k)
\right.
$$
\begin{equation}
\left.
+{N-1\over N}\sum_{j=1}^N \left(\frac{1}{3}-B_2(C_j)
\right) B_2(C_j) -{4\over 3}
\sum_{j,k=1}^N \left(B_3(C_j)-B_3(C_k)\right) B_1(C_{jk})
+{5\over144}(N^2-1)\right] \; ,
\label{twolooppot}
\end{equation}
where we introduce the variable
\begin{equation}
C_{j k} = C_j - C_k \; ,
\end{equation}
and the Bernoulli functions
\begin{equation}
B_1(x)=x-{1\over 2} \;\; , \;\;
B_2(x)=x^2-x+{1\over 6} \;\; , \;\;
B_3(x)=x^3-{3\over 2}x^2+{1\over 2}x \; ; \quad
\end{equation}
it is useful to note that for real $x$, which is
all that is needed here, $B_1(-x) = -B_1(x)$.
When $q_j = 0$, this reduces to the $\sim g^2$ free energy
for $\mu , T \neq 0$ \cite{kapusta}.

At two loop order, quarks contribute to the effective potential
from a single diagram in which a gluon crosses a quark loop.
The first term in (\ref{twolooppot}), involving
$B_2(C_{jk})$, represents the case in which 
the two quark lines in the loop
carry different $C_j$; then the exchanged
gluon is off-diagonal, and $C_{jk} = C_j - C_k$ enters.  
(Notice that the quark chemical potential drops out of $C_{jk}$.)
In contrast, the terms in (\ref{twolooppot})
involving $B_2(C_j)$ arise when both quark lines in the loop
have the same $C_j$, so that the exchanged gluon is neutral with
respect to the background field. This is true for the gluon in Feynman
gauge.

The origin of the terms $\sim B_1$ and $B_3$ in 
(\ref{twolooppot}) is due to two sources. The first is the longitudinal,
and so gauge dependent, part of the gluon propagator, which contributes
when the phase of the Wilson line is nonzero.

The second source was noted by Belyaev \cite{belyaev}: 
what is physically relevant is not the effective
potential for constant $A_0$, but that for the phase of the Wilson line.
This is most 
efficiently computed using a constrained functional integral
approach, as discussed by Korthals Altes \cite{altes}.
The effective potentials coincide at one loop order, but 
at two loop order, a finite renormalization of the Wilson line
must be taken into account.  This introduces  again $B_1$ and $B_3$,:
 $B_3$, for example, is the derivative of the one loop potential.
The gauge dependence from this source cancels precisely against that of the
longitudinal part of the gluon propagator.

We draw two conclusions from our results.
Consider first the limit of zero temperature.
The parameters for the phase of the Wilson line, the $q_j$, enters into
the $\mu _j \sim q_j T$, (\ref{mudef}).  From 
the effective potential in terms of the $\mu _j$'s,
(\ref{oneloopmu}), at one loop order
the effective potential only depends upon the $q_j$ through terms
which are at least {\it quadratic} in the temperature, 
$\sim \mu_j^4 \sim q_j^2 \mu^2 T^2$.  This remains true at two loop order,
although it is less obvious for a potential
written in terms of the $C_j$'s, (\ref{twolooppot}), than in terms
of the $\mu _j$'s.

That the effective potential for the phase of the Wilson line is
flat at zero temperature is reasonable.
A phase for the Wilson line is parametrized by a constant $A_0$
field; at nonzero temperature this cannot 
be undone by a gauge transformation, because it alters the
boundary conditions at $T\neq 0$.  At zero temperature, however,
the gauge transformation required to undo a constant $A_0$ field
is at infinite time, and so inconsequential.

We stress that our result implies that only the {\it phase} of the
Wilson line is independent of $\mu $ at $T=0$.
For a theory with dynamical quarks, the Wilson line has a nonzero
vacuum expectation value at $\mu = T = 0$, with a phase which
vanishes.  When $T=0$, as $\mu $ increases the vacuum expectation
value of the Wilson line will as well, as the effects of baryon,
or equivalently quark, loops enter.  
This increase is not calculable from the 
present work.  What we do show is that nothing interesting happens
with the phase.  Remember that at $T \neq 0$ and $\mu =0$,
there are metastable states associated with the phase of the
Wilson line \cite{misc}.  Our results show that there is nothing
analogous when $T=0$ and $\mu \neq 0$.

Our second conclusion is when $T \neq 0$, 
any point at which a $q_j =\pm 1/2$ is special.  From 
the definition of $\mu _j$, (\ref{mudef}),
for such a $q_j$ the energy $k^0$ changes from an
odd multiple of $\pi T$ (plus the chemical potential), into an
even multiple.  The corresponding part of the free energy, ${\cal S}_j$,
is then that of a free boson at $\mu , T \neq 0$.
Thus whenever $q_j = \pm 1/2$, the quarks ``Bose condense'':
the true ground state is not a filled Fermi sea, as for any other
value of $q_j$, but a state in which there is a macroscopic density
of states with zero momentum.  
Depending upon the $q_j$'s, several colors may condense at the same
time.  For two colors, for example, one can have all colors
Bose condense at the same point.  When $q_j = \pm 1/2$, the two
loop terms in the effective potential are just some perturbative
corrections.  

This does not violate the spin-statistics theorem, because a system with
$q_j \neq 0$ does not satisfy standard thermodynamic properties.
For example, from (\ref{contour}), the statistical distribution function is 
$n(k)= 1/(exp((k - \mu _j)/T) + 1)$; when
$q_j = \pm 1/2$, $\mu _j = \mu + i \pi T$, and $n(k)$
is {\it minus} the statistical
distribution function for bosons!  
Because of behavior such as this, some authors have argued that
$q_j \neq 0$ is an artifact of the imaginary time approach \cite{anti}.
We do not think so (see, e.g, Holland and Wiese, and 
Bronoff et al. \cite{roberge}), 
but do not presume to have assuaged the doubts of others.

While we have only computed at one and two loop order, we believe that
our two principal results --- that the potential is flat for $\mu \neq 0$
and $T = 0$, and that there is
``Bose condensation'' for $\mu _j = \pm 1/2$ --- are
true order by order in the loop expansion.  In particular, 
in going from one to two loop order, there is no sign that the effective
potential develops any singularity as $T \rightarrow 0$.

The work of R.D.P. is supported in part by DOE grant
DE-AC02-98CH10886.

\end{document}